\documentclass[11pt]{article}
\usepackage{amsmath,amssymb,dsfont}
\newcommand{\beq}{\begin{equation}}
\newcommand{\eeq}{\end{equation}}
\newcommand{\ba}{\begin{array}}
\newcommand{\ea}{\end{array}}
\newcommand{\bea}{\begin{eqnarray}}
\newcommand{\eea}{\end{eqnarray}}
\newcommand{\bean}{\begin{eqnarray*}}
\newcommand{\eean}{\end{eqnarray*}}

\newtheorem{theorem}{Theorem}[section]
\newtheorem{conjecture}{Conjecture}

\newtheorem{proof}{Proof.}

\newcounter{appendix}
\newcommand{\CG}{{\cal G}}

\newcommand{\CL}{{\cal L}}

\def\ga{\gamma}         
\def\be{\beta}
\def\al{\alpha}
\def\ep{\epsilon}
\def\l{\lambda}
\def\la{\lambda}        
\def\de{\delta}         
         
\def\sig{\sigma}

\newcommand{\jmp}[3]{Jour. Math. Phys. {\bf #1} #2 (#3)
}

\newcommand{\half}{\frac{1}{2}}

\def\back{\!\!\!\!\!\!}

\def\mat2#1#2#3#4{{\left(\begin{array}{cc}#1 & #2\\ #3 & #4
      \end{array}\right)}}

\def\mats2#1#2#3#4{{\left(\begin{array}{cc}#1 & #2\vspace{2truemm} \\ #3 & #4
\end{array}\right)}}

\def\endpf{\begin{flushright}$\square$\end{flushright}}

\def\alg{{\mathfrak g}}

\begin{document}
\begin{titlepage}
\begin{center}
{\Huge Algebraic extensions of Gaudin models}
\end{center}
\vspace{0.8truecm}
\begin{center}
{\Large
Fabio Musso$^\diamondsuit$, Matteo Petrera$^\sharp$, Orlando Ragnisco$^\flat$
\vskip0.8truecm
Dipartimento di Fisica E. Amaldi \\
Universit\`a degli Studi di Roma Tre\\
and\\
Istituto Nazionale di Fisica Nucleare,
Sezione di Roma Tre\\
Via della Vasca Navale 84, 00146 Roma, Italy\\
\vspace{0.8truecm}
E--mail$^\diamondsuit$:  musso@fis.uniroma3.it\\
E--mail$^\sharp$:  petrera@fis.uniroma3.it\\
E--mail$^\flat$:  ragnisco@fis.uniroma3.it}
\end{center}
\vspace{0.2truecm}
\vspace{0.2truecm}
\abstract{\noindent

We perform a In\"on\"u--Wigner contraction on Gaudin models,
showing how the integrability property is preserved
by this algebraic procedure. Starting from Gaudin models we obtain new integrable
chains, that we call Lagrange chains, associated to the same linear $r$-matrix 
structure. We give a general construction involving
rational, trigonometric and elliptic solutions of the classical Yang-Baxter equation.
Two particular examples are explicitly considered: the rational Lagrange chain and the trigonometric one.
In both cases local variables of the models are the generators of the direct sum of 
$N$ $\mathfrak{e}(3)$ interacting tops.

}\vskip 2truecm \noindent
Keywords: classical Yang-Baxter equation, Gaudin models, algebraic contractions, spinning tops \\
PACS: 02.30Ik, 45.40.Cc.
\end{titlepage}


\section{Introduction}

In the present paper we are concerned with an extension of Gaudin model. Since it was
introduced in the 70's the Gaudin model \cite{Gaudin1,Gaudin2} 
attracted considerable interest among theoretical and
mathematical physicists, and played a distinguished role in the realm of integrable
systems. Its peculariar properties, holding both at the classical and at the quantum level,
are deeply connected with the long--range nature of the interaction described by its commuting 
Hamiltonians, which in fact yields a typical ``mean field'' dynamic.

The algebraic counterpart of these physical features lies in the linear $r$-matrix structure,
that characterizes both the classical and the quantum model, and holds whatever be
the dependence (rational, trigonometric, elliptic) upon the spectral parameter. The linear $r$-matrix
structure does not depend on  the site number $N$ (i.e. on the number of copies of the
underlying Lie algebra) and moreover is not affected by a In\"on\"u--Wigner contraction
procedure. 
Such robustness property can for instance unable one to inherit the integrability property of the
Lagrange top model from that of a two--site rational Gaudin magnet \cite{KPR}. 
More generally, it allows one to
construct, out of the $N$--site Gaudin model, novel interacting systems (both classical and quantum),
whose underlying algebraic structure has been denoted in the literature as 
``jet extensions'' of Lie algebras \cite{H,H2}.

In this paper we will elucidate the connection between Lie algebra contractions and Lie algebra 
jet extensions in a general and systematic fashion, and we will derive the associated families
of classical (and quantum) integrable models.

We have to mention here that pioneering results in this direction were obtained already in the 80's
by Faddeev and Reshetikhin in their seminal paper \cite{RF}, also, 
we have to mention the results obtained  by
Harnad in \cite{H2} for the rational case.
However, we want to stress the connection among the systems investigated in the present paper
and the Gaudin models, that to the best of our knowledge, is established here for the first time.

\section{Drinfel'd--Belavin solutions to cYBE and Gaudin Models}

Let $\cal{P}$ be a Poisson manifold whose symplectic leaves have generically dimension $N$. 
Differentiable functions on $\cal{P}$ form a Poisson algebra and a completely integrable 
system on $\cal{P}$ is specified giving an abelian Poisson subalgebra of dimension $N$.
In the modern theory of classical integrable systems the generators of this abelian Poisson
subalgebra $\{H_1,\dots,H_N\}$, called Hamiltonians of the system, are usually obtained as spectral 
invariants of a Lax matrix $L(\lambda)$ depending on a complex parameter $\lambda$ and 
whose matrix elements are functions on the Poisson manifold ${\cal{P}}$.
Let us denote with $M$ the dimension of the square matrix $L(\lambda)$ defined by
\beq
L(\l) = \sum_{ij} L_{ij}(\l) \, E_{ij}, \nonumber
\eeq
where $(E_{ij})_{kl} = \delta_{ik} \delta_{jl}$ is the canonical basis of the $ M \times M$ matrices.
Introducing $\mathds{1}$ as the identity matrix of dimension $M$ 
we define $\{ L(\la) \otimes \mathds{1}, \mathds{1} \otimes L(\mu) \}$ as the matrix of Poisson
brackets between the entries of $L(\l)$:
\beq
\{ L(\la) \otimes \mathds{1}, \mathds{1} \otimes L(\mu) \} \doteq
\sum_{ij,kl} \{ L_{ij}(\l), L_{kl}(\mu) \} \,E_{ij} \otimes E_{kl}. \nonumber
\eeq
In this context a theorem due to Babelon and Viallet \cite{BV} ensures that the commutativity
of the Hamiltonians $\{H_1,\dots,H_N\}$ is equivalent to the existence of a $M^2 \times M^2$ matrix 
$r_{12}(\la,\mu)$ such that:
\begin{equation}
\{ L(\la) \otimes \mathds{1}, \mathds{1} \otimes L(\mu) \}=[r_{12}(\la,\mu), 
 L(\la) \otimes \mathds{1}]-[r_{21}(\mu,\lambda),\mathds{1} \otimes L(\mu)], \label{BV1}
\end{equation}
where 
\beq
r_{21}(\lambda,\mu)=\Pi \,  r_{12}(\la,\mu)\, \Pi, \nonumber
\eeq
with $\Pi$ the permutation operator on $\mathbb{C}^M \otimes \mathbb{C}^M$:
\beq
\Pi(x \otimes y)= y \otimes x.\nonumber
\eeq

In this paper we will deal with an 
interesting subcase for equation (\ref{BV1})
obtained when one imposes the following constraints on $r_{12}$ and $r_{21}$:
\beq
r_{12}(\la,\mu)=r_{12}(\la-\mu)=-r_{21}(\la-\mu), \nonumber
\eeq
and moreover requires that the $r-$matrix be ``not dynamical'' (i.e. independent of
phase space variables).

In this case equation (\ref{BV1}) becomes (omitting the indices $12$):
\begin{equation}
\{ L(\la) \otimes \mathds{1}, \mathds{1} \otimes L(\mu) \}+[r(\la-\mu), 
 L(\la) \otimes \mathds{1}+\mathds{1} \otimes L(\mu)]=0 .\label{BV2}
\end{equation}
It is well known that, in the case of a constant $r$-matrix, a sufficient condition
for the r.h.s. of (\ref{BV2}) to give rise to a well defined Poisson bracket is that
$r$ satisfies the so-called ``classical Yang-Baxter equation'' (cYBE):
\begin{equation}
[r_{13}(\la),r_{23}(\mu)]+[r_{12}(\la-\mu),r_{13}(\la)+r_{23}(\mu)]=0 ,\label{cYB}
\end{equation}
where 
\beq
r_{12}=r \otimes \mathds{1}, \qquad r_{23}=\mathds{1} \otimes r, \qquad
r_{13}= ( \mathds{1} \otimes \Pi)\, r_{12}\, ( \mathds{1} \otimes \Pi). \nonumber
\eeq

This equation was studied by Drinfel'd and Belavin \cite{D} in the case when the $r-$matrix is an 
element of $\alg \otimes \alg$, with $\alg$ being a finite-dimensional simple Lie algebra.
Once one has such a solution $r(\la)$ of equation (\ref{cYB}), then $(\rho_1 \otimes \rho_2)\,r(\la)$
with $\rho_1$ and $\rho_2$ faithful representations for $\alg$, gives a matrix solution for 
the cYBE.
 
Let 
$\{X^\al\}_{\al=1}^{{\rm{dim}}\, \alg}$ be a basis of the Lie algebra $\alg$, which is defined by the
commutation relations
\beq
\left[X^\al, X^\be \right]=C^{\al \be}_\ga \, X^\ga, \nonumber
\eeq
and let $g_{\al \be}$ be the inverse of the Cartan--Killing metric related to this basis.
Drinfel'd proved that under suitable non-degeneracy conditions on $r(\la)$ the only solutions
to cYBE (\ref{cYB}) are given by:
\begin{itemize}
\item rational solution: $ \displaystyle r(\l)=\frac{g_{\al \be} \, X^\al \otimes X^\be}{\la} $,
\item trigonometric solution: $ \displaystyle r(\l) =\sum_{n=-\infty}^{\infty} \frac{(A^n \otimes \mathds{1}) \,
g_{\al \be} \, X^\al \otimes X^\be}{\lambda-n \omega}  $,
\item elliptic solution: $ \displaystyle  r(\l) =\sum_{n,m=-\infty}^{\infty} \frac{(A^n B^m\otimes \mathds{1}) \,
g_{\al \be} \, X^\al \otimes X^\be}{\lambda-n \omega_1-m \omega_2} \ , \qquad [A,B]=0.$ 
\end{itemize}
Here $A$ and $B$ denote two finite order commuting automorphisms of $\alg$ not having
a common fixed vector. Since $\mathfrak{sl}(n)$ is the only simple Lie algebra possessing 
such two automorphisms it follows that the elliptic solution can be defined only 
in the case $\alg=\mathfrak{sl}(n)$.

For our purposes, the relevant property of Drinfel'd solutions is that
all of them can be generically written in the form:
\begin{equation}
r(\la)=g_{\al \be} \, X^\al \otimes X^\be f^\al(\la), \label{Dsol}
\end{equation}
for suitable choices of the functions $f^\al(\la)$.

From the cYBE we obtain a system of functional equations
that must be satisfied by the functions $f^\al(\la)$. In fact the first member of (\ref{cYB}) 
is given by
\bea
&& \left[ g_{\al \be} \, X^\al \otimes 1 \otimes X^\be  \, f^\al(\la), 
g_{\ga \de} \, 1 \otimes X^\ga \otimes X^\de  \, f^\ga(\la)\right]= \nonumber \\
&& =g_{\al \be}\,  g_{\ga \de} \,C^{\be \de}_\eta \, X^\al \otimes X^\ga 
\otimes X^\eta \, f^\al(\la) \, f^\ga(\mu), \label{fm}
\eea
while from the second one we obtain
\begin{eqnarray}
&&\left[ g_{\al \be} \, X^\al \otimes X^\be \otimes 1 \, f^\al(\la-\mu), g_{\ga \eta} \,
(X^\ga \otimes 1 \otimes X^\eta  \, f^\ga(\la)+
1 \otimes X^\ga \otimes X^\eta \, f^\ga(\mu)) \right]= \nonumber \\
&& =  g_{\al \be} \, g_{\ga \eta} \, f^\al(\la-\mu)\, \left[ C^{\al \ga}_\de \, X^\de 
\otimes X^\be \otimes X^\eta \, f^\ga(\la)+
C^{\be \ga}_\de \, X^\al \otimes X^\de \otimes X^\eta \, f^\ga(\mu) \right]= \nonumber \\
&& = g_{\al \be} \, g_{\ga \de} \, X^\al \otimes X^\ga \otimes X^\eta \,
C^{\be \de}_\eta \, \left[ f^\de (\la-\mu) f^\be(\la)-
f^\al(\la -\mu) f^\de (\mu) \right], \label{sm}
\end{eqnarray} 
where the last member is obtained from the previous one using the identity
$
g_{\gamma \eta} \, C^{\alpha \gamma}_{\delta}= - g_{\gamma \delta} \,
C^{\alpha \gamma}_{\eta}
$
and swapping the indexes $\gamma \leftrightarrow \beta, \alpha
\leftrightarrow \delta$ in the first term and  $\gamma
\leftrightarrow \delta$ in the second one.

Equating (\ref{fm}) and (\ref{sm}) we obtain the system of functional equations:
\begin{equation}
\sum_{\be, \delta=1}^{{\rm{dim}}\,\alg} 
g_{\al \be} \, g_{\ga \de} \, C^{\be \de}_\eta \, 
\left[ f^\al(\la) f^\ga(\mu) + f^\de (\la-\mu) f^\be(\la)-
f^\al(\la -\mu) f^\de (\mu) \right]=0, \label{functional}
\end{equation}
for all $\al,\gamma,\eta =1,...,{\rm{dim}}\,\alg$.
Note that in  equation (\ref{functional}) the summation is not on all the repeated indices, so that
the sum is indicated explicitly.

To each matrix $r(\la)$ we can associate a Lax matrix whose entries are functions on the Lie--Poisson
manifold ${\cal{P}}_{\alg}$ associated with $\alg$ in the following way: 
we choose a faithful representation $\rho$ of $\alg$
and we denote by $\{y^\al\}_{\al=1}^{{\rm{dim}}\alg}$ the coordinate functions related to the basis dual
to $\{X^\al\}_{\al=1}^{{\rm{dim}}\,\alg}$:
\begin{displaymath}
\left[X^\al, X^\be \right]=C^{\al \be}_\ga \, X^\ga \qquad \Longrightarrow \qquad 
\left\{y^\al, y^\be\right\}=C^{\al \be}_\ga y^\ga.
\end{displaymath}    
We define our Lax matrix as:
\begin{equation}
L(\la)=\tau+g_{\al \be} \,  \rho(X^\al) \, y^\be \, f^\al(\la), \label{L1}
\end{equation}
where $\tau \in \rho(\alg)$ is a constant matrix chosen in such a way that 
\beq
[\tau \otimes \mathds{1} + \mathds{1} \otimes \tau, r(\l) ]=0 . \nonumber
\eeq
This Lax matrix satisfies the linear $r$-matrix algebra
\begin{displaymath}
\{ L(\la) \otimes \mathds{1}, \mathds{1} \otimes L(\mu) \}+[r(\la-\mu), 
 L(\la) \otimes \mathds{1}+\mathds{1} \otimes L(\mu)]=0,
\end{displaymath}
with respect to the ordinary Lie--Poisson bracket, where $r$ is given by (\ref{Dsol}).

Now let us consider the Lie--Poisson manifold associated with the direct sum of $N$ copies 
of a simple Lie algebra $\alg$. We denote with $\{y^\al_i\}_{\al=1}^{{\rm{dim}}\,\alg}$
the coordinates on the dual of the $i-$th copy of $\alg$:
\begin{equation}
\left\{ y^\al_i, y^\be_j \right\}= \delta_{ij} \, C^{\al \be}_\ga \, y^\ga_i .\label{locality}
\end{equation}   
On each site we can define a matrix $L_i(\la)$:
\begin{equation}
L_i(\la)=g_{\al \be} \, \rho(X^\al)\, y^\be_i \, f^\al(\la-\la_i), \label{Li}
\end{equation}
where for any $i$ we have shifted the spectral parameter by a fixed quantity $\lambda_i \in \mathbb{C}$.
Each matrix $L_i$ will obviously satisfy equation (\ref{BV2}) and thanks to the locality of the Poisson 
relations (\ref{locality}) so will do also the total Lax matrix:
\begin{equation}
L_\CG(\la)=\tau+\sum_{i=1}^N L_i(\la). \label{LG}
\end{equation}
It follows that the spectral invariants of the Lax matrix $L_\CG(\la)$ will provide us
with a set of involutive Hamiltonians that define the $N-$sites {\it{Gaudin model}}
associated to $\alg$ \cite{Gaudin1,Gaudin2,Sk89,Jurco,RSTS}.

\section{Algebraic contractions of Gaudin models}

In the present section we illustrate our main result. Starting from Gaudin Lax operators (\ref{LG})
we obtain, through a In\"on\"u--Wigner contraction, new Lax matrices satisfying the same
$r$-matrix structure pertaining to the starting ones.

Let us recall the notion of generalized In\"on\"u--Wigner contraction \cite{IW}. Let $V$ a complex finite,
or infinite dimensional vector space. Let $\mathfrak{h}=(V,\mu)$ be a Lie algebra with Lie multiplication
$\mu: V \times V \rightarrow V$. 

The analytic notion of contraction can be described by a continuous family of homomorphism
\beq
U(\epsilon): V \rightarrow V, \quad \epsilon \in (0,1], \quad U(1)=1, \nonumber
\eeq
which are nonsingular for $\epsilon >0$ and singular for $\epsilon=0$.

The new Lie bracket on $V$,
\beq
\mu_\epsilon(a,b)=U(\epsilon)\, \mu (U^{-1}(\ep)\,a, U^{-1}(\ep)\,b), \quad a,b \in V, \quad \ep>0,
\nonumber
\eeq
corresponds to a change of basis given by $U(\ep)$, and leads to the Lie algebra
$\mathfrak{h}_\ep=(V,\mu_\ep)$ isomorphic to $\mathfrak{h}$. If
\beq
\mu' (a,b) =\lim_{\ep \rightarrow 0} \mu_\epsilon(a,b)\nonumber
\eeq
exist for all $a,b \in V$, we call $\mathfrak{h}'=(V,\mu')$ the contraction of $\mathfrak{h}$ by $U(\ep)$;
$\mathfrak{h}'$ is, in general, not isomorphic to $\mathfrak{h}$.

For a generalized In\"on\"u--Wigner contraction, $U(\ep)$ has a simple form, namely
\beq
V= \bigoplus_{i=0}^N V^{(i)}, \quad N \geq 1, \nonumber
\eeq 
\beq
U(\ep) |_{V^{(i)}} = \ep^{n_i} \mathds{1} |_{V^{(i)}}, \label{con}
\eeq
with
\beq
0 \leq n_0 <n_1 < ... < n_M, \quad n_i \in \mathbb{R}, \quad i=0,...,M.\nonumber
\eeq
The necessary and sufficient condition for (\ref{con}) to define a contraction is
\beq
\mu \left(V^{(i)},V^{(j)} \right) \subset \bigoplus_l V^{(l)},\label{fabio}
\eeq
with
\beq
n_l \leq n_i + n_j,\nonumber
\eeq
where all surviving structure constants are the same as for $\mathfrak{h}$. The special case $n_0=0$
and $N=1$ is called a simple In\"on\"u--Wigner contraction.

We will apply a generalized In\"on\"u--Wigner contraction to the direct sum of $N$ copies
of a simple Lie algebra $\mathfrak{h}$:
\beq
\mathfrak{h} = \overbrace{\alg \oplus \dots \oplus \alg}^N.\nonumber
\eeq
The contraction procedure will be carried out directly on the Lie--Poisson algebra associated with
$\mathfrak{h}$. We denote again with
$\{y^\al_i\}_{\al=1}^{{\rm{dim}}\,\alg}$
the coordinates on the dual of the $i-$th copy of $\alg$, satisfying
$
\{ y^\al_i, y^\be_j \}= \delta_{ij} \, C^{\al \be}_\ga \, y^\ga_i
$ as in formula (\ref{locality}).

Since we need (\ref{fabio}) we make the following change of basis in $\mathfrak{h}$: 
\beq
\left(
\begin{array}{c}
\tilde y_0^\al\\
\tilde y_1^\al\\
\vdots \\
\tilde y_{N-1}^\al
\end{array} \right) =
\left(
\begin{array}{ccccc}
1 & 1 & \dots & 1 \\
\nu_1 &  \nu_2& \dots &  \nu_{N} \\
\vdots & \vdots & \ddots & \vdots \\
 \nu_1^{N-1}&  \nu_2^{N-1} & \dots & 
\nu_{N}^{N-1}
\end{array} \right)
\left(
\begin{array}{c}
y_1^\al\\
y_2^\al\\
\vdots \\
y_N^\al
\end{array} \right), \qquad \al=1,...,{\rm{dim}}\,\alg, \nonumber
\eeq 
where the matrix defining
the change of basis is a nonsingular Vandermonde matrix
since we take $\nu_{j} \neq \nu_{k}$, $1 \leq j \neq k \leq N$.

Note that in the new coordinates the equation
(\ref{fabio}) is satisfied if one identifies $V^{(i)}$ with the space generated by
$\{\tilde y^\al_i\}_{\al=1}^{{\rm{dim}}\,\alg}$ and $n_i$ with $i$.

We define the homomorphism $U(\ep)$ by
\beq
U(\ep) |_{V^{(i)}} = \ep^{i} \mathds{1} |_{V^{(i)}} \qquad i=1,...,N-1. \nonumber
\eeq
We absorb the above homomorphism $U(\ep)$ into the definition of the new coordinates
$\{\tilde y^\al_i\}_{\al=1}^{{\rm{dim}}\,\alg}$, namely hereafter we will denote
\beq
\tilde y^\beta_i \doteq \sum_{j=1}^N \ep^i \nu^i_j y^\be_j. \nonumber
\eeq
The new bracket $\mu_\ep$ is defined by
\bea 
\back \left\{ \tilde{y}^\al_i, \tilde{y}^\be_j \right\}_\ep &=& \sum_{k,m=1}^N \ep^{i+j} \, 
\nu_{k}^{i} \,  \nu_{m}^{j} \,  \left\{ y^\al_i, y^\be_j \right\} = \nonumber \\
&=&
C^{\al
\be}_\gamma \, \sum_{k=1}^N  \ep^{i+j} \, \nu_{k}^{i+j } \, y^\gamma_k= \nonumber \\
&=& \left\{ \begin{array}{cc}
C^{\al \be}_\gamma \, \tilde{y}^\gamma_{i+j} \quad &  i+j <N,\\
O(\ep) \quad & i+j \geq N.
\end{array} \right.
\nonumber \eea 
After contraction we obtain the new Lie--Poisson algebra $\mathfrak{h}'$:
\beq
\left\{ \tilde y^\al_i, \tilde y^\be_j \right\}= \left\{
\begin{array}{cc}
C^{\al \be}_\gamma \, \tilde y^\gamma_{i+j} & \quad i+j < N,\\
0 & \quad i+j \geq N.
\end{array} \right. \nonumber
\eeq

Let us give some remarks on the algebraic structure of our algebra $\mathfrak{h}'$. If we define
the new vector spaces
\bea
&& W^{(i)}= \bigoplus_{j=i}^{N-1} V^{(j)} \qquad i=0,...,N-1, \nonumber \\
&& W^{(0)} \simeq \mathfrak{h}' \nonumber,
\eea
we obtain a flag of ideals
\beq
\left\{ W^{(i)}, W^{(j)} \right\} \subseteq W^{(j)} \qquad j \geq i. \nonumber
\eeq
In particular 
\beq
\left\{\mathfrak{h}' , W^{(1)} \right\} \subseteq W^{(1)}. \nonumber
\eeq
So we have an exact sequence:
\beq
0 \longrightarrow W^{(1)} \stackrel{i}{\longrightarrow} \mathfrak{h}'
\stackrel{\pi}{\longrightarrow} \mathfrak{g} \longrightarrow 0 \nonumber
\eeq 
where $i$ is the inclusion and $\pi$ is the projection
on the quotient $\mathfrak{h}' / W^{(1)} $; moreover we use the fact that
$\mathfrak{h}' / W^{(1)} \simeq V^{(0)} \simeq \alg$.

We can therefore see the contraction procedure as an extension of the starting Lie algebra
$\alg$ that is known as the ``$N-$th jet extension'' of $\alg$ \cite{H}.

In the following theorem we give a recipe to extend the
 contraction procedure at the level of the equation (\ref{BV2}).

\begin{theorem} Let $\CL(\l)$ be the Lax matrix on the Lie--Poisson algebra $\mathfrak{h}'$
obtained from  $L_\CG(\l)$ (\ref{LG}) through the following procedure:
\begin{itemize}
\item make the identification $\l_i=- \epsilon \nu_{i}$,
\item take a formal power series expansion in $\ep$,
\item take the limit $\ep \rightarrow 0$. 
\end{itemize}
Then $\CL(\l)$ satisfies 
\beq
\left\{ \CL(\la) \otimes \mathds{1},  \mathds{1} \otimes \CL(\mu) \right\} + \left[ r(\la-\mu), 
\CL(\la)
\otimes \mathds{1} + \mathds{1} \otimes \CL(\mu) \right]=0, \label{linear}
\eeq
with respect to the Lie--Poisson algebra defined by
\beq
\left\{ \tilde y^\al_i, \tilde y^\be_j \right\}= \left\{
\begin{array}{cc}
C^{\al \be}_\gamma \, \tilde y^\gamma_{i+j} & \quad i+j < N,\\
0 & \quad i+j \geq N,
\end{array} \right. 
\label{jet}
\eeq
with the same $r$-matrix given by (\ref{Dsol}).
\end{theorem}

\noindent{\bf Proof:} 
Setting $\l_i=- \epsilon \nu_{i}$ in (\ref{LG}) we
easily obtain the formal expansion
\beq
L_\CG(\l; \ep) =\tau+ \sum_{i=1}^N  g_{\al \be} \, \rho(X^\al) \, y^\be_i\,  \sum_{j=0}^\infty 
\frac{(\ep \nu_i)^j}{j!} \left( \frac{d}{d \l} \right)^j f^\al (\l). \nonumber
\eeq
Through the identification
\beq
\tilde y^\beta_i \doteq \sum_{j=1}^N \ep^i \nu^i_j y^\be_j, \nonumber
\eeq 
we immediately get
\beq
L_\CG(\l; \ep) \sim
\tau + \sum_{i=0}^{N-1} \ell_i(\l) \, f^\al_i(\la)+ O(\ep), \nonumber
\eeq
where $\ell_i(\l)$ and $f^\al_i(\la)$ are given respectively by
\beq
\ell_i(\l) = g_{\al \beta}\, \rho(X^\al)\,   \tilde y^\be_i.\nonumber
\eeq 
\beq
f_{i}^\al(\l) = \frac{1}{i!}\,  \left(\frac{d}{d \l} \right)^i \,  
f_0^\alpha(\l), \qquad f^\al_0(\la) \doteq f^\al(\la) 
\label{Fi}
\eeq
Finally taking the limit $\ep \rightarrow 0$ we obtain:
\beq
\CL (\l) = \lim_{\ep \rightarrow 0} 
L_\CG(\l; \ep) =
\tau + \sum_{i=0}^{N-1} \ell_i(\l) \, f^\al_i(\la). \nonumber
\eeq
Let us prove (\ref{linear}). We set $\tilde{y}^\eta_{i+j}=0$ if $i+j>N$. We have:
\begin{eqnarray}
\left\{ \CL(\la) \otimes \mathds{1}, \mathds{1} \otimes \CL(\mu) \right\} &=& 
g_{\al \beta} \, g_{\gamma \delta} \, \rho(X^\al) \otimes
\rho(X^\gamma) \sum_{i,j=0}^{N-1} f^\al_i(\la) \, f^\gamma_j(\mu)
\left\{ \tilde{y}^\be_i, \tilde{y}^\delta_j \right\}= \nonumber \\
& =& g_{\al \beta} \, g_{\gamma \delta} \, \rho(X^\al) \otimes
\rho(X^\gamma) \, C^{\beta \delta}_{\eta} \sum_{i,j=0}^{N-1}
f^\al_i(\la) \, f^\ga_j(\mu) \, \tilde{y}^\eta_{i+j}= \nonumber \\
&=& g_{\al \beta} \, g_{\gamma \delta} \, \rho(X^\al) \otimes
\rho(X^\gamma) \, C^{\beta \delta}_{\eta} \sum_{i=0}^{N-1} \tilde{y}^\eta_i
\sum_{j=0}^i f^\al_j(\la) \, f^\ga_{i-j}(\mu) .\label{y}
\end{eqnarray}
On the other hand we have:
\begin{eqnarray}
&& \back \left[ r(\la-\mu), \CL(\la) \otimes \mathds{1} +  \mathds{1} \otimes \CL(\mu) \right]= 
g_{\al \beta} \, g_{\gamma \eta}\,  \times \nonumber \\
&&  \back \times
\left[  \rho(X^\al) \otimes \rho(X^\beta) \, f^\al(\l-\mu), 
\left( \rho(X^\gamma)
\otimes \mathds{1}\right) \sum_{i=0}^{N-1} \tilde y^\eta_i \,  f^\gamma_i (\l)
+\left(\mathds{1} \otimes \rho(X^\gamma) \right) \sum_{i=0}^{N-1}
\tilde y^\eta_i  \,  f_i^\gamma (\mu)
  \right] = \nonumber \\
&& \back =g_{\al \beta}  \,  g_{\gamma \eta} \,  f^\al(\l-\mu) \times \nonumber \\
&& \back \times
\left[
C^{\al \gamma}_\delta \,  
 \rho(X^\delta) \otimes \rho(X^\beta) \sum_{i=0}^{N-1}
\tilde y^\eta_i  \,  f^\gamma_i (\l)  +  C^{\beta \gamma}_\delta \,  
 \rho(X^\alpha) \otimes \rho(X^\delta) \sum_{i=0}^{N-1} \tilde y^\eta_i  \,  f_i^\gamma (\mu)
 \right]=\nonumber \\
&& \back = 
g_{\al \beta} \,  g_{\gamma \delta} \,
\rho(X^\al) \otimes \rho(X^\gamma) \, C^{\beta \delta}_{\eta}
\sum_{i=0}^{N-1} \tilde y^{\eta}_i \left[f^\delta(\l-\mu) f^\be_i(\l)- f^\al(\l-\mu)f^\de_i(\mu)  \right],
\label{yy}
\end{eqnarray}
where the last expression is obtained from the previous one using the identity
$
g_{\gamma \eta} \, C^{\alpha \gamma}_{\delta}= - g_{\gamma \delta} \,
C^{\alpha \gamma}_{\eta}
$
and swapping the indexes $\gamma \leftrightarrow \beta, \alpha
\leftrightarrow \delta$ in the first term and  $\gamma
\leftrightarrow \delta$ in the second one.

Considering equations (\ref{y}) and (\ref{yy}) we obtain
\bea
&&\left\{\CL(\la) \otimes \mathds{1},  \mathds{1} \otimes \CL(\mu) \right\} + \left[ r(\la-\mu), 
\CL(\la)
\otimes \mathds{1} + \mathds{1} \otimes \CL(\mu) \right]=g_{\al \beta} \, g_{\gamma \delta} \, 
\rho(X^\al) \otimes \rho(X^\gamma) \,\times \nonumber \\
&& \quad \qquad \times \, C^{\beta \delta}_{\eta}
\sum_{i=0}^{N-1} \tilde y^\eta_i
\left[ \sum_{j=0}^i f^\al_j(\la) f^\ga_{i-j}(\mu) +f^\delta(\l-\mu) f^\be_i(\l)- f^\al(\l-\mu)f^\de_i(\mu)
\right]. \label{21}
\eea
Let us look in more detail at the quantity enclosed in the square brackets
in equation (\ref{21}).
Using (\ref{Fi}) we have:
\bea
\sum_{j=0}^i f^\al_j(\la) \, f^\ga_{i-j}(\mu) &=&
\sum_{j=0}^i \frac{1}{j!} \left( \frac{d}{d \la} \right)^j f^\al(\la) \frac{1}{(i-j)!} 
\left( \frac{d}{d \mu} \right)^{i-j} f^\ga(\mu)= \nonumber \\
&=& \frac{1}{i!} \sum_{j=0}^i {i \choose j}\left( \frac{d}{d \la} \right)^j
\left( \frac{d}{d \mu} \right)^{i-j} f^\al(\la) \, f^\ga(\mu)= \nonumber \\
&=& \frac{1}{i!} \left( \frac{d}{d\l}+ \frac{d}{d\mu}\right)^i f^\al(\la) \, f^\ga(\mu),
\label{po}
\eea
and 
\bea 
f^\delta(\l-\mu) f^\be_i(\l)&-& f^\al(\l-\mu)f^\de_i(\mu) = \nonumber \\
& =& \frac{1}{i!} \left[
f^\delta(\l-\mu) \left( \frac{d}{d \la} \right)^i f^\be(\l)
- f^\al(\l-\mu) \left( \frac{d}{d \mu} \right)^i f^\de(\mu) \right]= \nonumber \\
&=&\frac{1}{i!} \left( \frac{d}{d\l}+ \frac{d}{d\mu}\right)^i
\left[ f^\de (\la-\mu) f^\be(\la)-
f^\al(\la -\mu) f^\de (\mu) \right].
\label{po2}
\eea
Finally, from (\ref{po}) and (\ref{po2}) we obtain
\begin{eqnarray*}
&&\back \left\{\CL(\la) \otimes \mathds{1},  \mathds{1} \otimes \CL(\mu) \right\} + \left[ r(\la-\mu), 
\CL(\la)
\otimes \mathds{1} + \mathds{1} \otimes \CL(\mu) \right]= g_{\al \beta} \, g_{\gamma \delta} \, 
\rho(X^\al) \otimes \rho(X^\gamma) \, C^{\beta \delta}_{\eta} \times \nonumber \\
&&\back \quad \times 
\sum_{i=0}^{N-1} \tilde y^\eta_i \, 
\frac{1}{i!} \left( \frac{d}{d\l}+ \frac{d}{d\mu}\right)^i
\left[ f^\al(\la) f^\ga(\mu) + f^\de (\la-\mu) f^\be(\la)-
f^\al(\la -\mu) f^\de (\mu) \right]=\\
&&\back =\sum_{i=0}^{N-1} \tilde y^\eta_i \, \frac{1}{i!}\left( \frac{d}{d\l}+ \frac{d}{d\mu}\right)^i
\times \nonumber \\
&& \times \left\{ g_{\al \beta} \, g_{\gamma \delta} \, \rho(X^\al) \otimes \rho(X^\gamma) \, C^{\beta \delta}_{\eta}
\left[ f^\al(\la) f^\ga(\mu) + f^\de (\la-\mu) f^\be(\la)-
f^\al(\la -\mu) f^\de (\mu) \right] \right\},
\end{eqnarray*}
that vanishes thanks to equation (\ref{functional}).
The fact that (\ref{jet}) is a Lie--Poisson algebra is then a plain consequence of 
(\ref{linear}).
\endpf

\section{Extensions of $\mathfrak{su}(2)$  Gaudin models: Lagrange chains}

In this section we will consider algebraic contractions of $\mathfrak{su}(2)$  Gaudin models.
We choose the matrices $\sigma^1,\sigma^2,\sigma^3$ as a basis of the 
fundamental representation of $\mathfrak{su}(2)$:
$$
\sigma^1 \doteq  \frac{{\rm{i}} \, \sigma_x}{2} = \half \left(\begin{array}{cc}
0 & {\rm{i}}  \\
{\rm{i}} & 0
\end{array}\right), \quad
\sigma^2 \doteq  \frac{{\rm{i}} \, \sigma_y}{2} = \half \left(\begin{array}{cc}
0 & 1  \\
-1 & 0
\end{array}\right), \quad
\sigma^3 \doteq \frac{{\rm{i}} \, \sigma_z }{2}= \half \left(\begin{array}{cc}
{\rm{i}} & 0  \\
0 & -{\rm{i}}
\end{array}\right),
$$
where $\sigma_x,\sigma_y,\sigma_z$ are the Pauli matrices.
In this basis the 
$r-$matrix (\ref{Dsol}) can be written in the form:
\beq
r(\la)=- \sum_{\al=1}^3 f_0^{\al}(\l) \,  \sig^{\al} \otimes \sig^{\al} \label{RR}
\eeq 
where
\begin{equation}
f_0^1(\l)=\left\{ \begin{array}{cc}
\frac{1}{\l} & \rm{rational ~~case}\\
\frac{1}{\sin \l } & \qquad ~~  \rm{trigonometric ~~case}, \\
\frac{\rm{dn}}{\rm{sn}} (\l)    & \rm{elliptic~~ case},
\end{array} \right. \label{uno}
\end{equation}
\begin{equation}
f_0^2(\l)=\left\{ \begin{array}{cc}
\frac{1}{\l} & \rm{rational ~~case}\\
\frac{1}{\sin \l }& \qquad ~~  \rm{trigonometric ~~case}, \\
\frac{1}{\rm{sn}} (\l)    & \rm{elliptic~~ case},
\end{array} \right. \label{due}
\end{equation}
\begin{equation}
f_0^3(\l)=\left\{ \begin{array}{cc}
\frac{1}{\l} & \rm{rational ~~case}\\
\cot \l & \qquad ~~ \rm{trigonometric ~~case}, \\
\frac{\rm{cn}}{\rm{sn}} (\l)   & \rm{elliptic~~ case}.
\end{array} \right. \label{tre}
\end{equation}

Let us consider the Lax matrix of the $\mathfrak{su}(2)$ Gaudin magnet:
\beq
L_\CG(\la)=\tau+  \sum_{i=1}^N \sum_{\al=1}^3
\sigma^\al x^\al_i \, f^\al_0 (\la-\la_i), \label{jj}
\eeq
where $\tau$ can be chosen as:
\begin{itemize}
\item any $\mathfrak{su}(2)$-matrix in the rational case,
\item $\sigma^3$ in the trigonometric case,
\item the identity in the elliptic case,
\end{itemize}  
and the local variables of the model $\{ x^\al_i \}$, $\al=1,2,3$, $i=1,...,N$ are the generators 
of the direct sum of $N$ $\mathfrak{su}(2)$ spins with the
following Lie--Poisson brackets:
\beq
\left\{ x^\al_i, x^\be_j \right\} = \delta_{ij} \, x^\gamma_i \qquad i,j=1,...,N \nonumber
\eeq
where $\al \be \gamma$ is the cyclic permutation of 123.

The first natural extension of Gaudin models is obtained by 
contracting two copies of the Lie--Poisson algebra
$\mathfrak{su}(2)$ ($N=2$ in the Lax matrix (\ref{jj})), namely
\beq
\mathfrak{su}(2) \oplus \mathfrak{su}(2) \simeq \mathfrak{o}(4) \rightarrow \mathfrak{e}(3)
\nonumber
\eeq
where $\mathfrak{e}(3)$ is the real euclidean Lie--Poisson algebra in the 3-space. The Lie--Poisson
brackets of $\mathfrak{e}(3)$ are given by
\beq
\left\{    y^{\alpha},y^{\beta}   \right\}=  y^{\gamma}, \qquad
\left\{    y^{\alpha},z^{\beta}   \right\}=  z^{\gamma}, \qquad
\left\{    z^{\al}, z^{\be}  \right\}= 0 , \qquad \alpha,\beta,\gamma= 1,2,3. \nonumber
\eeq
After contraction we get the new Lax matrix:
\beq
L (\l) = \tau  + \sum_{\alpha=1}^3
\sig^{\alpha} \left[f^{\al}_0 (\l)\, y^{\al}
+  f^{\al}_1 (\l) \, z^{\al}  \right], \label{mat}
\eeq
where $f^{\al}_0 (\l)$ is given by (\ref{uno}),(\ref{due}),(\ref{tre}) and 
$ f^{\al}_1 (\l)$ is given by (\ref{Fi}).
Note that in the rational case the Lax matrix (\ref{mat}) is exactly
the Lax matrix of the Lagrange top \cite{Au,RSTS}. This particular contraction procedure
on the two--site $\mathfrak{su}(2)$ Gaudin magnet has been considered in \cite{KPR}.

Moreover we observe that the spectral invariants of the Lax matrix (\ref{mat}) are
the same for the rational, trigonometric and elliptic cases. However 
this is no more true if we
extend our Lax matrix (\ref{mat}) to the 
$N$-bodies case as we did for Gaudin model (see equations (\ref{L1}),(\ref{Li}),(\ref{LG})).
Namely we consider the 
Lax matrix associated to the Lie--Poisson algebra of the direct sum of $N$ copies of
$\mathfrak{e}(3)$:
\beq
\CL(\l)= \tau +  \sum_{i=1}^N   \sum_{\alpha=1}^3
\sig^{\alpha} \left[f^{\al}_0 (\l - \l_i)\, y_i^{\al}
+  f^{\al}_1 (\l - \l_i)\, z_i^{\al}  \right]. \nonumber
\eeq

\subsection{The rational Lagrange chain}

The rational $\bigoplus_{i=1}^N \mathfrak{e}_i(3)$ chain, that we call
rational Lagrange chain is derived from the following $2 \times 2$ Lax matrix
\beq
\CL(\l)= w \sigma^3 +  \sum_{i=1}^N   \sum_{\alpha=1}^3
\sig^{\alpha} \left[\frac{y_i^{\al}}{\l - \l_i}
+  \frac{z_i^{\al}}{(\l - \l_i)^2 } \right],\label{laxrational}
\eeq
where $\l_i \in \mathbb{C}$ and $w \in \mathbb{R}$ are the parameters of the model
and $\l \in \mathbb{C}$ is the spectral parameter. The parameter 
$w$ denotes the intensity of an external field, taken as uniform (along the chain)
and constant (in time).

Local variables of the model are generators of the direct sum of $N$ $\mathfrak{e}(3)$ tops,
$y_i^{\al}, z_i^{\al}$, $i=1,...,N$, $\al=1,2,3$ with the following Lie--Poisson brackets:
\beq
\left\{    y_i^{\alpha},y_j^{\beta}   \right\}= \delta_{ij} \, y_j^{\gamma}, \qquad
\left\{    y_i^{\alpha},z_j^{\beta}   \right\}= \delta_{ij} \, z_j^{\gamma}, \qquad
\left\{    z_i^{\al},z_j^{\be}   \right\}= 0 , \label{e30}
\eeq
where $\al \be \gamma$ is the cyclic permutation of 123.

The Lie--Poisson brackets (\ref{e30}) have $2N$ Casimir functions:
\beq
C_i^{(1)} = \sum_{\alpha=1}^3 y^\al_i z^\al_i, \qquad
C_i^{(2)} = \sum_{\alpha=1}^3 (z^\al_i)^2. \nonumber
\eeq 

As we have shown in the previous section the
 Lax matrix (\ref{laxrational}) satisfies the linear $r$-matrix algebra
\beq
\left\{\CL(\l) \otimes \mathds{1}, \mathds{1} \otimes \CL(\mu) \right\}+
\left[r(\l - \mu), \CL(\l) \otimes \mathds{1} + \mathds{1} \otimes \CL(\mu) \right]=0,
\label{333}
\eeq
where $\mathds{1}$ is the $2 \times 2$ identity matrix and the $r$-matrix is given by
\beq
r(\la)=-\frac{1}{\l} \sum_{\al=1}^3  \sig^{\al} \otimes \sig^{\al}.
\label{666}
\eeq
The spectral curve $\Gamma$,
\beq
\Gamma: \quad \det( \CL(\l) -\mu \mathds{1})=0, \nonumber
\eeq
is an hyperelliptic curve of genus $g=2N-1$, given by
\beq
-\mu^2 =
w^2+
\sum_{i=1}^N \frac{R_i}{\l-\l_i}+
\frac{S_i}{(\l-\l_i)^2}+
\frac{C_i^{(1)}}{(\l-\l_i)^3}+
\frac{C_i^{(2)}}{(\l-\l_i)^4},\nonumber
\eeq 
with the Hamiltonians $R_i$ and $S_i$ given by:
\begin{eqnarray}
R_i &=& w y^3_i+ \sum_{k \neq i }^N
\left(
\frac{y^1_i y^1_k + y^2_i y^2_k + y^3_i y^3_k}{\l_i - \l_k}+ \right. \nonumber \\
& & + \left. \frac{y^1_i z^1_k + y^2_i z^2_k + y^3_i z^3_k - y^1_k z^1_i - y^2_k z^2_i - y^3_k z^3_i}
{(\l_i - \l_k)^2}
 - 2 \frac{z^1_i z^1_k + z^2_i z^2_k + z^3_i z^3_k}{(\l_i - \l_k)^3}
\right), \nonumber \\
&& \nonumber \\
S_i &=& \half \left[(y^1_i)^2+(y^2_i)^2+ (y^3_i)^2 \right]  + w  z^3_i+ \nonumber \\
&& +
\sum_{k \neq i }^N
\left(
\frac{y^1_k z^1_i + y^2_k z^2_i + y^3_k z^3_i}{\l_i - \l_k}+
 \frac{z^1_i z^1_k + z^2_i z^2_k + z^3_i z^3_k}{(\l_i - \l_k)^2}
\right). \nonumber 
\end{eqnarray}
These are integrals of motion of the rational Lagrange chain, 
which are Poisson commuting:
\beq
\left\{R_i,R_j\right\}=\left\{S_i,S_j\right\}=\left\{R_i,S_j\right\}=0, \quad i,j=1,...,N. \nonumber
\eeq
Notice that there is one linear integral given by
\beq
\sum_{i=1}^N R_i= w \sum_{i=1}^N y^3_i. \nonumber
\eeq
Let us remark that if $N=1$ the Lagrange chain degenerates into the well-known 
$\mathfrak{e}(3)$ Lagrange top, whose Hamiltonians are given by
$y^3$ and $(y^1)^2+(y^2)^2+(y^3)^2 + 2 w  z^3$.

We can bring the curve $\Gamma$ into the canonical form by the scaling
\beq
\mu \longmapsto \hat \mu = \mu \prod_{i=1}^{N} (\l -\l_i)^2.\nonumber
\eeq
The equation of the spectral curve becomes
\bea
- \hat \mu^2 & =& \left[
w^2+
\sum_{i=1}^N \frac{R_i}{\l-\l_i}+
\frac{S_i}{(\l-\l_i)^2}+
\frac{C_i^{(1)}}{(\l-\l_i)^3}+
\frac{C_i^{(2)}}{(\l-\l_i)^4}  \right] \, \prod_{i=1}^{N} (\l -\l_i)^4 = \nonumber \\
&=& w^2 \l^{4N}+s_1 \l^{4N-1} +s_2 \l^{4N-2} + ...+ s_{4N},
\eea
where the coefficients $s_j$, $j=1,...,4N$ are linear combinations of the Hamiltonians
and of the Casimir functions.

\subsection{The trigonometric Lagrange chain}

The trigonometric $\bigoplus_{i=1}^N \mathfrak{e}_i(3)$ chain, that we call
trigonometric Lagrange chain, is derived from the following $2 \times 2$ Lax matrix
\beq
\CL(\l)= w \sigma^3 +  \sum_{i=1}^N   \sum_{\alpha=1}^3
\sig^{\al} \left[ ~ f_0^{\al}(\l-\l_i) ~ y_i^{\al}
+  f_1^{\al}(\l-\l_i) ~ z_i^{\al} \right], \label{Lt}
\eeq
where
\begin{eqnarray}
&& f_0^1 (\l) = \frac{1}{\sin \l}, \qquad
f_0^2(\l) = \frac{1}{\sin \l}, \qquad
f_0^3 (\l) = \cot \l, \nonumber \\
&& f_1^1 (\l) = \frac{\cot \l}{\sin \l}, \qquad
f_1^2 (\l) = \frac{\cot \l}{\sin \l}, \qquad
f_1^3 (\l) = \cot^2 \l +1. \nonumber
\end{eqnarray}
The
Lax matrix (\ref{Lt}) satisfies the linear $r$-matrix algebra (\ref{333}) with $r$-matrix given
by (\ref{RR},\ref{due}).

The spectral curve $\Gamma$,
\beq
\Gamma: \quad \det( \CL(\l) -\mu \mathds{1})=0, \nonumber
\eeq 
is given by
\beq
-\mu^2 =
Q+
\sum_{i=1}^N \cot(\l-\l_i)R_i+
\cot^2(\l-\l_i)\, S_i+
\cot^3(\l-\l_i)\, C_i^{(1)}+
\cot^4(\l-\l_i)\, C_i^{(2)}, \nonumber
\eeq 
where
\begin{eqnarray}
Q &=& w^2 + \sum_{i=1}^N \left[2 w z^3_i + (y^1_i)^2+(y^2_i)^2 \right] +\nonumber \\
&& + 2 \sum_{i=1}^N\,\sum_{k \neq i }^N \left\{ \frac{1}{\sin(\l_i-\l_k)} \left[
 y^1_k z^1_i + y^2_k z^2_i + \cos(\l_i-\l_k) y^3_k z^3_i
\right]+ \right.\nonumber \\
&& \left. + \frac{2}{\sin^2(\l_i-\l_k)} \left[
\cos(\l_i-\l_k) (z^1_k z^1_i + z^2_k z^2_i) + z^3_k z^3_i
\right]  - y^3_i y^3_k \right\}, \nonumber \\
&& \nonumber 
\end{eqnarray}
\begin{eqnarray}
R_i &=& 2 w y^3_i+ (z^1_i)^2+(z^2_i)^2+(z^3_i)^2+  \nonumber \\
&& +2  \sum_{k \neq i }^N \left\{ \frac{1}{\sin(\l_i-\l_k)} \left[\frac{}{}
 y^1_k y^1_i + y^2_k y^2_i + \cos(\l_i-\l_k) y^3_k y^3_i - \right. \right.  \nonumber \\
&& \left. \left. - \left(\frac{2}{\sin^2(\l_i-\l_k)} -1
\right) (z^1_k z^1_i + z^2_k z^2_i) - \frac{2\cos(\l_i-\l_k)}
{\sin^2(\l_i-\l_k)} z^3_k z^3_i
\right] \right. + \nonumber \\
&& \left. + \frac{1}{\sin^2(\l_i-\l_k)} \left[
 \cos(\l_i-\l_k) (y^1_i z^1_k + y^2_i z^2_k - y^1_k z^1_i - y^2_k z^2_i) + y^3_i z^3_k - y^3_k z^3_i
\right] \right\}, \nonumber \\
&& \nonumber \\
S_i &=& (y^1_i)^2+(y^2_i)^2+(y^3_i)^2 + 2 w z^3_i  + (z^1_i)^2+(z^2_i)^2+ (z^3_i)^2 + \nonumber \\&& + 2 \sum_{k \neq i }^N \left\{ \frac{1}{\sin(\l_i-\l_k)} \left[
 y^1_k z^1_i + y^2_k z^2_i + \cos(\l_i-\l_k) y^3_k z^3_i
\right]+ \right.\nonumber \\
&& \left. + \frac{1}{\sin^2(\l_i-\l_k)} \left[
\cos(\l_i-\l_k) (z^1_k z^1_i + z^2_k z^2_i) + z^3_k z^3_i
\right] \right\}. \nonumber 
\end{eqnarray}
Note that the coefficient $Q$ is functional dependent on the $S_i$, $R_i$ and $C_i^{(2)}$:
\beq
Q= w^2+ \sum_{i=1}^N S_i - \left(\sum_{i=1}^N \frac{R_i}{2w}  \right)^2 +
\left(\sum_{i=1}^N \frac{C_i^{(2)}}{2w}  \right)^2. \nonumber
\eeq 

As in the rational case we identify in the coefficients $R_i$, $S_i$ the Poisson commuting 
integrals of the trigonometric Lagrange chain:
\beq
\left\{R_i,R_j\right\}=\left\{S_i,S_j\right\}=\left\{R_i,S_j\right\}=0, \quad i,j=1,...,N.\nonumber
\eeq 

\section{Concluding remarks}

In this paper we have performed a In\"on\"u--Wigner contraction on Gaudin models,
showing that the integrability property is preserved
by this algebraic procedure. Other interesting structures can be inherited from the Gaudin models 
as well. For example in \cite{KPR} it is shown how to obtain B\"acklund
transformations for the Lagrange top from those of the $\mathfrak{su}(2)$ Gaudin model. 
It turns out that it
is possible to generalize this approach to the construction of B\"acklund
transformations for Lagrange chains.
This work is in progress and we plan to report the results in a forthcoming paper.


Finally we want to remark that the contraction procedure described in this paper can be applied
to quantum Gaudin models as well, in a straightforward way.

\section*{Acknowledgments}

The authors wish to acknowledge the support from the GNFM (Gruppo Nazionale per la Fisica
Matematica), research project: ``Onde nonlineari, struttura tau e geometria delle
variet\`a invarianti: il caso della gerarchia di Camassa-Holm''.

\section*{Appendix: The general rational Lax matrix}

In the proof of the theorem 3.1 we have seen that our Lax matrix $\CL (\l)$ satisfies the 
$r$-matrix structure
\beq
\left\{\CL(\la) \otimes \mathds{1},  \mathds{1} \otimes \CL(\mu) \right\} + \left[ r(\la-\mu), 
\CL(\la)
\otimes \mathds{1} + \mathds{1} \otimes \CL(\mu) \right]=0\nonumber
\eeq
provided that the functions $f^\al_j(\la)$ satisfy the following system of functional equations 
(see equation (\ref{21})):
\beq
\sum_{\beta,\delta=1}^{{\rm{dim}}\, \alg}
g_{\al \beta} \, g_{\gamma \delta} \, C^{\beta \delta}_{\eta} \,
\left[ \sum_{j=0}^i f^\al_j(\la) f^\ga_{i-j}(\mu) +f^\delta(\l-\mu) f^\be_i(\l)- f^\al(\l-\mu)f^\de_i(\mu)
\right]=0, \label{ew}
\eeq
for all $\al, \be,\ga$ and $i=0,\dots,N-1$.

In theorem 3.1 we have shown that the system (\ref{ew}) is satisfied with the choice (\ref{Fi}); however 
this is not the general analytic solution to the system (\ref{ew}).
In this appendix we give a conjecture about the general analytic solution  of (\ref{ew})
in the rational case, where we have
\beq
f^\al_0(\la)= \frac{1}{\l - c_0},\nonumber
\eeq 
$c_0$ being an arbitrary constant; $f^\al_0(\la)$ obviously does not depend on $\al$.

The system of functional equations (\ref{ew}) reduces to the following one:
\begin{equation}
\sum_{i=0}^k f_i(\la)
f_{k-i}(\mu)=\frac{f_k(\la)-f_k(\mu)}{\mu-\la} \qquad k=0,\dots,N.
\label{sr}
\end{equation}
Our solution is based on the following algebraic conjecture:
\begin{conjecture}
Let $k \in \mathbb{N}$, $c_1,\dots,c_k$ arbitrary constants and
let us denote with $\{\vec{J}_i\}$ the set of vectors satisfying
the Diophantine equation:
\begin{displaymath}
\{\vec{J}_i\}=\{ \vec{J} \in \mathbb{N}^i \ : \ J_1+2 J_2+\cdots+i
J_i=i \},
\end{displaymath}
then the following polynomial identity holds:
\beq
\sum_{i=0}^k \left( \sum_{\{\vec{J}_i\} }\frac{c_1^{J_1}}{J_1!}
\dots \frac{c_i^{J_i}}{J_i!} \lambda^{|\vec{J}_i|} \sum_{ \{
\vec{J}_{k-i} \} } \frac{c_1^{J_1}}{J_1!}\dots
\frac{c_{k-i}^{J_{k-i}}}{J_{k-i}!} \mu^{|\vec{J}_{k-i}|} \right)
=\sum_{\{\vec{J}_k\} } \frac{c_1^{J_1}}{J_1!} \dots
\frac{c_k^{J_k}}{J_k!} (\lambda+\mu)^{|\vec{J}_k|}, \nonumber
\eeq
where we denoted with $|\vec{J}_i|$ the $1-$norm of $\vec{J}_i$:
\begin{displaymath}
|\vec{J}_i|=\sum_{l=1}^i J_l.
\end{displaymath}
\end{conjecture}
Using the program for symbolic manipulation Maple we tested this
conjecture for $N \leq 20$.

Using the explicit form for the function $f_0(\l)$ and our conjecture
it is not difficult to construct the general analytic solution for
the system (\ref{sr}), reading:
\begin{equation}
f_i(\lambda)=\sum_{\{\vec{J}_i\} } \frac{c_1^{J_1}}{J_1!} \dots
\frac{c_i^{J_i}}{J_i!} \left( \frac{d}{d \lambda}
\right)^{|\vec{J}_i|} f_0(\lambda) \qquad i=1,\dots,N.
\label{solfi}
\end{equation}
In fact, inserting (\ref{solfi}) into (\ref{sr}) we get:
\begin{eqnarray*}
&& \sum_{i=0}^k f_i(\la)
f_{k-i}(\mu)=\\
&&=\sum_{i=0}^k \left( \sum_{\{\vec{J}_i\} }
\frac{c_1^{J_1}}{J_1!} \dots \frac{c_i^{J_i}}{J_i!} \left(
\frac{d}{d \lambda} \right)^{|\vec{J}_i|} f_0(\lambda)
\sum_{\{\vec{J}_{k-i}\} } \frac{c_1^{J_1}}{J_1!} \dots
\frac{c_{k-i}^{J_{k-i}}}{J_{k-i}!} \left( \frac{d}{d \mu}
\right)^{|\vec{J}_{k-i}|} f_0(\mu) \right)=\\
&&= \sum_{\{\vec{J}_k\} } \frac{c_1^{J_1}}{J_1!} \dots
\frac{c_k^{J_k}}{J_k!} \left( \frac{d}{d \lambda} + \frac{d}{d
\mu} \right)^{|\vec{J}_k|} \left(f_0(\lambda) f_0(\mu) \right)=\\
&&=\sum_{\{\vec{J}_k\} } \frac{c_1^{J_1}}{J_1!} \dots
\frac{c_k^{J_k}}{J_k!} \left( \frac{d}{d \lambda} + \frac{d}{d
\mu} \right)^{|\vec{J}_k|} \left(\frac{f_0(\la)-f_0(\mu)}{\mu-\la} \right) =\\
 &&= \frac{f_k(\la)-f_k(\mu)}{\mu-\la}.
\end{eqnarray*}
The fact that solutions (\ref{solfi}) provide the general
analytic solution to (\ref{sr}) can be argued taking the limit
$\lambda \rightarrow \mu$ in the functional equations
(\ref{sr}), yielding the system of ordinary differential
equations:
\begin{equation}
\sum_{i=0}^k f_i(\la) f_{k-i}(\la)=-\frac{d f_k(\la)}{d \la}
\qquad k=0,\dots,N .\label{trsystem}
\end{equation}
As the system (\ref{trsystem}) is triangular,
for any given $i$ 
the functions $f_i(\la)$ can be found solving a system
of $i+1$ ordinary differential equations and therefore depend at
most upon $i+1$ arbitrary parameters. This is exactly the number
of arbitrary parameters entering our solutions (\ref{solfi}).

Using our conjecture it is straightforward to extend the
solutions (\ref{solfi}) to the trigonometric and elliptic case. It
is enough to replace $f_i(\la)$ with $f^\alpha_i(\la)$ and
$f_0(\la)$ with the corresponding $f_0^\alpha(\la)$ given by
(\ref{uno},\ref{due},\ref{tre}):
\begin{displaymath}
f^\alpha_i(\lambda)=\sum_{\{\vec{J}_i\} } \frac{c_1^{J_1}}{J_1!}
\dots \frac{c_i^{J_i}}{J_i!} \left( \frac{d}{d \lambda}
\right)^{|\vec{J}_i|} f^\alpha_0(\lambda) \qquad i=1,\dots,N.
\end{displaymath}

\vfill

\pagebreak

\label{lastpage}

\end{document}